\documentclass[12pt]{article}

\usepackage{amsmath,amssymb,amsfonts,amsbsy}
\usepackage{cite}
\usepackage{graphicx}
\usepackage{wrapfig}
\usepackage[font=small,labelfont=bf,labelsep=period]{caption}

\begin{document}
\addcontentsline{toc}{subsection}{{Title of the article}\\
{\it B.B. Author-Speaker}}

%%%%%%% please do not touch these! %%%%%%
\setcounter{section}{0}
\setcounter{subsection}{0}
\setcounter{equation}{0}
\setcounter{figure}{0}
\setcounter{footnote}{0}
\setcounter{table}{0}

%%% Version with comments of Ernst Oct. 31,  TO ADD a paragraph on transverse polarization Oct 29, 2009

\begin{center}
\textbf{LONGITUDINAL SPIN TRANSFER TO $\Lambda$ AND $\bar{\Lambda}$ IN POLARIZED PROTON-PROTON COLLISIONS AT $\sqrt{s} = 200\,\mathrm{GeV}$}

\vspace{5mm}

Qinghua Xu$^{\dag}$ for the STAR Collaboration \\

\vspace{2mm}

{\it Department of Physics, Shandong University, Shandong 250100, China} 

\vspace{4mm}

\begin{small}
  $\dag$ \emph{E-mail: xuqh@sdu.edu.cn}
\end{small}
\end{center}

\vspace{0.0mm} % Don't laugh: it does change the spacing!

\begin{abstract}
  We report our measurement on longitudinal spin transfer, $D_{LL}$, from high energy polarized protons to $\Lambda$ and $\bar{\Lambda}$ hyperons in proton-proton collisions at $\sqrt{s} = 200\,\mathrm{GeV}$ with the STAR detector at RHIC.
The measurements cover $\Lambda$, $\bar\Lambda$ pseudorapidity $\left|\eta\right| < 1.2$ and transverse momenta, $p_\mathrm{T}$, up to $4\,\mathrm{GeV}/c$.
The longitudinal spin transfer is found to be $D_{LL}= -0.03\pm 0.13(\mathrm{stat}) \pm 0.04(\mathrm{syst})$
 for inclusive $\Lambda$ and $D_{LL} = -0.12 \pm 0.08(\mathrm{stat}) \pm 0.03(\mathrm{syst})$ for inclusive $\bar{\Lambda}$ hyperons with  $\left<\eta\right> = 0.5$ and $\left<p_\mathrm{T}\right> = 3.7\,\mathrm{GeV}/c$.
The $p_T$ dependence of $D_{LL}$ for positive and negative $\eta$ is given. 
\end{abstract}

\vspace{7.2mm} 

%$\Lambda$ hyperon has been studied extensively in different aspects of spin physics, due to their self-analyzing weak decay.
The longitudinal polarization of $\Lambda$ hyperons has been studied in $e^+e^-$ annihilation and lepton-nucleon deep inelastic scattering (DIS) with polarized beams and/or targets.
Such polarization studies provide access to polarized fragmentation function and the spin content of $\Lambda$ \cite{aleph}.
Here we report the first measurement on longitudinal spin transfer ($D_{LL}$) from the proton beam to $\Lambda ({\bar \Lambda})$ produced in proton-proton collisions at $\sqrt s$=200 GeV\cite{STAR_DLL},
\begin{equation}
D_{LL}\equiv \frac
{\sigma_{p^+p \to  \Lambda ^+ X}-\sigma_{p^+p \to  \Lambda ^-X}}
{\sigma_{p^+p \to  \Lambda ^+ X}+\sigma_{p^+p \to  \Lambda ^-X}},
\label{gener1}
\end{equation}
where the superscript $+$ or $-$ denotes the helicity.
Within the factorization framework, the production cross sections are described in terms of calculable partonic cross sections and non-perturbative parton distribution and fragmentation functions.
The production cross section has been measured for transverse momenta, $p_\mathrm{T}$, up to about $5\,\mathrm{GeV}/c$ and is well described by perturbative QCD evaluations~\cite{Abelev:2006cs}.
The spin transfer $D_{LL}$ is thus expected to be sensitive to polarized fragmentation function and helicity distribution function of nucleon, as reflected in different model predictions of $D_{LL}$ at RHIC~\cite{deFlorian:1998ba,Boros:2000ex,Ma:2001na,Xu:2002hz}.

The spin transfer $D_{LL}$ in Eq.~(\ref{gener1}) is equal to the polarization of $\Lambda$ $(\bar{\Lambda})$ hyperons, ${P}_{\Lambda(\bar\Lambda)}$, if the proton beam is fully polarized.
 ${P}_{\Lambda(\bar\Lambda)}$ can be measured via the weak decay channel $\Lambda \to p \pi^-$ $(\bar \Lambda \to \bar p \pi^+)$ from the angular distribution of the final state,
\begin{equation}
\frac{\mathrm{d}N}{\mathrm{d} \cos{\theta}^*}=\frac{
\sigma\,\mathcal{L}\,A}{2}
(1+\alpha_{\Lambda(\bar\Lambda)} P_{\Lambda(\bar\Lambda)}
\cos{\theta}^*),
\label{ideal}
\end{equation}
where %$N_\mathrm{tot}$ is the total number of acceptance corrected $\Lambda$ $(\bar\Lambda)$'s, 
$\sigma$ is the (differential) production cross section, $\mathcal{L}$ is the integrated luminosity, $A$ is the detector acceptance, which is in general a function of $\cos\theta^*$ as well as other observables, $\alpha_{\Lambda}$=$-\alpha_{\bar{\Lambda}} = 0.642 \pm 0.013$~\cite{Amsler:2008zz} is the weak decay parameter, and 
$\theta^*$ is the angle between the $\Lambda$($\bar\Lambda$)  polarization direction and the \mbox{(anti-)proton} momentum in the $\Lambda$ ($\bar\Lambda$) rest frame.

The data presented here were collected at the Relativistic Heavy Ion  Collider (RHIC) with the Solenoidal Tracker at RHIC (STAR)~\cite{Ackermann:2002ad} in the year 2005.
An integrated luminosity of 2\,$\mathrm{pb}^{-1}$ was sampled with average longitudinal beam polarization of $52 \pm 3\%$ and $48 \pm 3\%$ for two beams.
The proton polarization was measured for each beam and each beam fill using Coulomb-Nuclear Interference (CNI) proton-carbon polarimeters~\cite{Jinnouchi:2004up}.
Different beam spin configurations were used for successive beam bunches and the pattern was changed between beam fills.
The data were sorted by beam spin configuration.

The analyzed data sample includes three different triggers. 
One is the minimum bias (MB) trigger sample, defined with a coincidence signal
from Beam-Beam Counters (BBC) at both sides of STAR interaction region.
The other data samples were recorded with the MB trigger condition and with two additional trigger conditions: a high-tower (HT) and a jet-patch (JP).
The HT trigger condition required the BBC proton collision signal in coincidence with a transverse energy deposit $E_\mathrm{T} > 2.6\,\mathrm{GeV}$ in at least one Barrel Electromagnetic Calorimeter (BEMC) tower, covering $\Delta\eta \times \Delta\phi=0.05 \times 0.05$ in pseudorapidity $\eta$ and azimuthal angle $\phi$.
The JP trigger condition imposed the MB condition in coincidence with an energy deposit $E_\mathrm{T} > 6.5\, \mathrm{GeV}$ in at least one of six BEMC patches each covering $\Delta\eta \times \Delta\phi=1 \times 1$.
The total BEMC coverage was $0 < \eta < 1$ and $0 < \phi < 2\pi$ in 2005.

%
%\begin{figure}
%\begin{center}
% \includegraphics[height=.22\textheight]{figure-1.eps}
 \begin{wrapfigure}[17]{R}{95mm}%% [number of text lines to wrap]{horizontal position: LRC}{figure width}
 \centering
  \vspace*{-8mm} %% the vertical position may need tweaking
  \includegraphics[width=95mm]{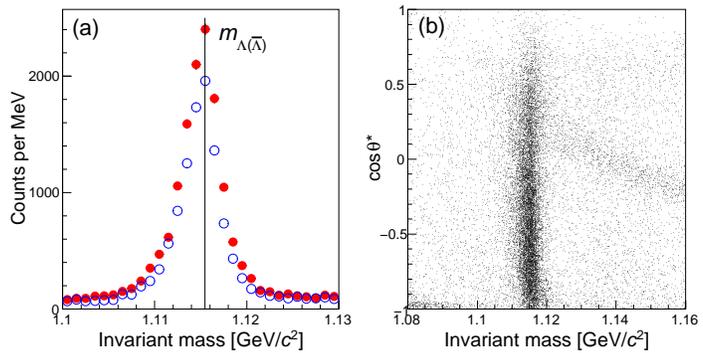}
 \caption{(a) The invariant mass distribution of $\Lambda$ (filled circles) and $\bar \Lambda$ (open circles) candidates from reconstructed $p + \pi^-$ and $\bar p + \pi^+$ track pairs in 2005 MB data after topological selections.
(b) The invariant mass distribution versus $\cos\theta^*$ for $\Lambda$.}
\label{mass}
%\end{center}
\end{wrapfigure}

The  $\Lambda$ ($\bar{\Lambda}$) candidates were identified from the topology of the weak decay channel to $p \pi^-$ ($\bar p \pi^+$), which has a branching ratio of 63.9\%~\cite{Amsler:2008zz}. 
Charged particle tracks in the 0.5\,T magnetic field were measured with the Time Projection Chamber (TPC), covering $0 < \phi < 2\pi$ and $|\eta|\lesssim 1.3$.
The charge tracks after particle identification from specific energy loss $dE/dx$ in TPC were paired to form a $\Lambda (\bar{\Lambda})$ candidate and topological selections were applied to reduce background\cite{STAR_DLL,Abelev:2006cs}.
The selections included criteria for the distance of closest approach between the paired tracks and the distance between the point of closest approach and the beam collision vertex, and demanded that the momentum sum of the track pair pointed at the collision vertex.
The criteria were tuned to preserve the signal while reducing the background fraction to 10\% or less.

Figure~\ref{mass}(a) shows the invariant mass distribution for the $\Lambda$ (filled circles) and $\bar\Lambda$ (open circles) candidates reconstructed from MB data with $|\eta| < 1.2$ and $0.3 < p_\mathrm{T} < 3\,\mathrm{GeV}/c$.
The mean values of the $\Lambda$ and $\bar{\Lambda}$ mass distributions are in agreement with the PDG mass value $m_{\Lambda(\bar\Lambda)} = 1.11568\,\mathrm{GeV}/c^2$~\cite{Amsler:2008zz}.
Figure~\ref{mass}(b) shows the same invariant mass distribution versus $\cos\theta^*$ for the $\Lambda$ candidates.
The number of $\Lambda$ candidates varies with $\cos\theta^*$ because of detector acceptance.
The small variation of the reconstructed invariant mass with $\cos\theta^*$ is understood to originate from detector resolution.
In addition to signal, combinatorial background is seen as well as backgrounds of misidentified $e^+e^-$ pairs at low invariant mass values near $\cos\theta^*=-1.0$ and of misidentified $K_{S}^0$ in a diagonal band at high invariant mass values and $\cos\theta^*>-0.2$.
About $1.2 \times 10^4\ \Lambda$ and $1.0 \times 10^4\ \bar{\Lambda}$ candidates from the MB sample were reconstructed in the invariant mass range $1.109< m <1.121\,\mathrm{GeV}/c^2$ and were kept for further analysis.

To minimize the uncertainty associated with acceptance, the longitudinal spin transfer $D_{LL}$ was extracted in small $\cos\theta^*$ intervals as follows:
\begin{equation}D_{LL}=\frac{1}{\alpha P_\mathrm{beam} \left<\cos \theta^*\right>} \frac{N^+ -R N^-} {N^+ + R N^-},\label{eq_dll}
\end{equation}
where $P_{beam}$ is the beam polarization, $N^+$ ($N^-$) are the $\Lambda$($\bar\Lambda$) counts in a small $\cos\theta^*$ interval when the beam is positively (negatively) polarized, $\left<\cos\theta^*\right>$ is the average value in this $\cos\theta^*$ bin, and $R=L^+/L^-$ is the corresponding luminosity ratio for these two polarization states.
Eq.(3) uses the parity conservation in the hyperon production in $pp$ collisions, which leads to a sign flip of hyperon polarization when the beam helicity is flipped.
The detector acceptance in a $\cos\theta^*$ interval is taken as a constant 
and thus canceled.
For this measurement, only one polarized beam is needed.
At RHIC both beams are polarized and the single spin yields $N^+$ ($N^-$) are thus formed from the yields $n^{++}$, $n^{+-}$, $n^{-+}$, and $n^{--}$ by beam helicity configuration weighted with their corresponding relative luminosities.
The luminosity ratios were measured with BBC at STAR \cite{Kiryluk:2005gg}.

 \begin{wrapfigure}[29]{R}{75mm}%% [number of text lines to wrap]{horizontal position: LRC}{figure width}
 \centering
  \vspace*{-8mm} %% the vertical position may need tweaking
  \includegraphics[width=75mm]{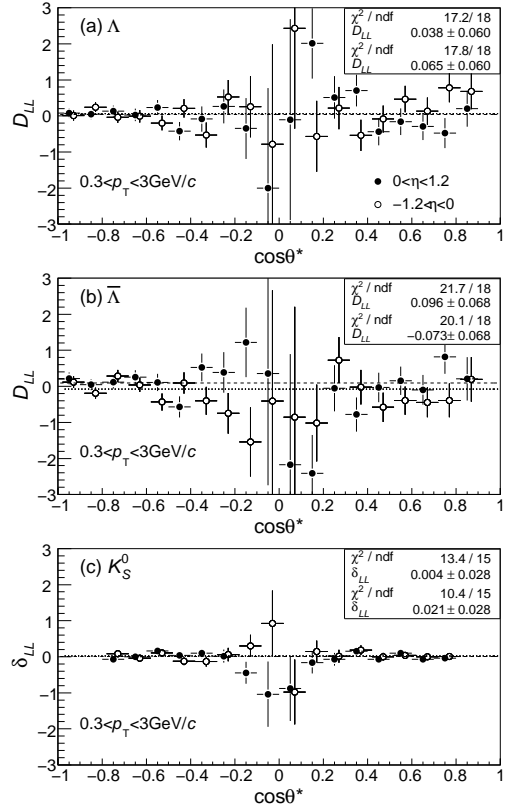}
   \caption{ The spin transfer $D_{LL}$ versus $\cos\theta^*$ for (a) $\Lambda$ and (b) $\bar \Lambda$, and (c) the spin asymmetry $\delta_{LL}$ for the control sample of $K^0_S$ mesons versus $\cos\theta^*$. 
%The filled circles show the results for positive pseudorapidities $\eta$ with respect to the polarized beam and the open circles show the results for negative $\eta$. 
Only statistical uncertainties are shown.
The data points with negative $\eta$ have been shifted slightly in $\cos\theta^*$ for clarity.
%The indicated values of $\chi^2$ and the spin transfer are for the data with positive and negative $\eta$, respectively.
}
\label{cosin_fit}
\end{wrapfigure}

The yields $N^+$ and $N^-$ were determined for each $\cos\theta^*$ interval from the observed $\Lambda$ and $\bar{\Lambda}$ candidate yields in the mass interval from 1.109 to 1.121\,GeV/$c^2$. 
The corresponding raw values $D^\mathrm{raw}_{LL}$ were averaged over 20 intervals covering the entire $\cos\theta^*$ range.
The obtained $D^\mathrm{raw}_{LL}$ values and their statistical uncertainties were then corrected for (unpolarized) background dilution according to $D_{LL} = D^\mathrm{raw}_{LL}/(1-r)$, where $r$ is the average background fraction.
No significant spin transfer asymmetry was observed for the yields in the sideband mass intervals $1.094 < m < 1.103\,\mathrm{GeV}/c^2$ and $1.127 < m < 1.136\,\mathrm{GeV}/c^2$, and thus no further correction was applied to $D_{LL}$.
However, a contribution was included in the estimated systematic uncertainty of the $D_{LL}$ measurement to account for the possibility that the background could nevertheless be polarized.

The combined $D_{LL}$ results from the MB data sample versus $\cos\theta^*$ are shown in Fig. \ref{cosin_fit}(a) for $\Lambda$ and Fig.\ref{cosin_fit}(b) for $\bar\Lambda$ hyperons with $0.3 < p_\mathrm{T} < 3\,\mathrm{GeV}/c$ and $0 < \eta < 1.2$ and $ -1.2 < \eta < 0$.
Positive $\eta$ is defined along the direction of the incident polarized beam.
Fewer than 50 counts were observed for $\cos\theta^* > 0.9$ and this interval was discarded for this reason.
The extracted $D_{LL}$ is constant with $\cos\theta^*$, as expected and confirmed by the quality of fit.
In addition, a null-measurement was performed of the spin transfer for the spinless $K_S^0$ meson, which has a similar event topology.
The $K_S^0$ candidate yields for $|\cos\theta^*| > 0.8$ were discarded since they have sizable $\Lambda (\bar{\Lambda}$) backgrounds.
The result, $\delta_{LL}$, obtained with an artificial weak decay parameter $\alpha_{K_S^0} = 1$, was found consistent with no spin transfer, as shown in Fig.~\ref{cosin_fit}(c).
%The analysis was furthermore tested with simulated $\Lambda$ data having a non-zero $D_{LL}$ and the $D_{LL}$ input to the simulation was extracted successfully.

The HT and JP data samples were recorded with trigger conditions that required large energy deposits in the BEMC, in addition to the MB condition.
These triggers, however, did not require a highly energetic $\Lambda$ or $\bar{\Lambda}$.
To minimize the effects of this bias, the HT event sample was restricted to $\Lambda$ or $\bar{\Lambda}$ candidates whose decay (anti-)proton track intersected a BEMC tower that fulfilled the trigger condition.
About 50\% of the $\bar{\Lambda}$ and only 3\% of the $\Lambda$ candidate events in the analysis pass this selection.
This is qualitatively consistent with the annihilation of anti-protons in the BEMC.
The $\bar\Lambda$ sample that was selected in this way thus directly triggered the experiment read-out.  
It contains about 1.0$\times 10^4$ $\bar\Lambda$ candidates with $1 < p_\mathrm{T} < 5\,\mathrm{GeV}/c$.
% and a residual background of about 5\%.

For the JP triggered sample, events were selected with at least one reconstructed jet that pointed to a triggered jet patch.
The same jet reconstruction was used as in Ref.~\cite{Abelev:2006uq}.
Jets outside the BEMC acceptance were rejected.
The $\Lambda$ and $\bar{\Lambda}$ candidates  whose reconstructed $\eta$ and $\phi$ fell within the jet cone of radius $r_\mathrm{cone} = \sqrt{(\Delta\eta)^2 + (\Delta\phi)^2} = 0.4$ were retained for further analysis.
In most cases, the decay (anti-)proton track is part of the reconstructed jet, whereas the decay pion track is not.
No correction was made to the jet finding and reconstruction for this effect.
About $1.3\times10^4~\Lambda$ and $2.1\times10^4~\bar{\Lambda}$ candidates with $1 < p_\mathrm{T} < 5\,\mathrm{GeV}/c$ remain after selections.
%The residual background is estimated to be 13\% for $\Lambda$ and 9\% for $\bar{\Lambda}$ candidates.

\begin{figure}
\begin{center}
 \includegraphics[height=.35\textheight]{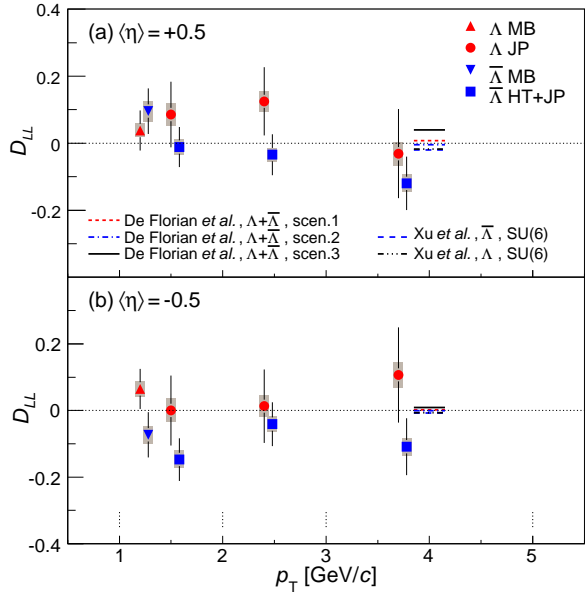}
\caption{
(Color online) Comparison of $\Lambda$ and $\bar \Lambda$ spin transfer $D_{LL}$ in  
polarized proton-proton collisions at $\sqrt{s} = 200$ GeV for (a) positive and  
(b) negative $\eta$ versus $p_\mathrm{T}$.  The vertical bars and bands indicate the  sizes of the statistical and systematic uncertainties, respectively.
The $\bar\Lambda$ data points have been shifted slightly in $p_\mathrm{T}$ for clarity.
The dotted vertical lines indicate the $p_\mathrm{T}$ intervals in the analysis of HT and JP data.
The horizontal lines show model predictions.
% evaluated at $\eta$ and largest $p_\mathrm{T}$ of the data.
}
\label{DLL_05eta}
\end{center}
\end{figure}

Figure~\ref{DLL_05eta} shows the $D_{LL}$ results of $\Lambda$ and $\bar{\Lambda}$ versus $p_\mathrm{T}$ for positive and negative $\eta$ respectively.
The $\bar{\Lambda}$ results from HT and JP data have been combined.
No corrections have been applied for possible decay contributions from heavier baryonic states.
The $\Lambda$ and $\bar{\Lambda}$ results for $D_{LL}$ are consistent with each other and consistent with no spin transfer from the polarized proton beam to the produced $\Lambda$ and $\bar{\Lambda}$ to within the present uncertainties.
The data have $p_\mathrm{T}$ up to $4\,\mathrm{GeV}/c$, where $D_{LL}= -0.03\pm 0.13(\mathrm{stat}) \pm 0.04(\mathrm{syst})$ for the $\Lambda$ and $D_{LL} = -0.12 \pm 0.08(\mathrm{stat}) \pm 0.03(\mathrm{syst})$ for the $\bar{\Lambda}$ at $\left<\eta\right> = 0.5$.
For reference, the model predictions of Refs.~\cite{deFlorian:1998ba,Xu:2002hz}, evaluated at $\eta=\pm 0.5$ and $p_\mathrm{T}=4\,\mathrm{GeV}/c$, are shown as horizontal lines.
The expectations of Ref.~\cite{deFlorian:1998ba} hold for $\Lambda$ and $\bar{\Lambda}$ combined and examine different polarized fragmentation scenarios, in which the strange (anti-)quark carries all or only part of the $\Lambda$ $(\bar{\Lambda})$ spin.
The model in Ref.~\cite{Xu:2002hz} separates $\Lambda$ from $\bar{\Lambda}$ and otherwise distinguishes the direct production of the $\Lambda$ and $\bar{\Lambda}$ from the (anti-)quark in the hard scattering and the indirect production via decay of heavier (anti-)hyperons.
%Both sets of expectations assume that the contribution from intrinsic gluon polarization can be neglected.
The evaluations are consistent with the present data and span a range of values that, for positive $\eta$, is similar to the experimental uncertainties.
The values for negative $\eta$ are expected to be negligible and thus less sensitive~\cite{deFlorian:1998ba,Xu:2002hz}.
The current experimental uncertainties are statistics limited, and we expect future data be able to distinguish between different models of polarized fragmentation and parton distribution functions.

The total systematic uncertainty in $D_{LL}$ is varying from 0.02 to 0.04 with increasing $p_\mathrm{T}$ and is smaller than the statistical uncertainty, ranging from 0.06 to 0.14, for each of the data points.
In estimating the size of the systematic uncertainties, we have combined contributions from the uncertainties in decay parameter $\alpha$ and in the measurements of the proton beam polarization and relative luminosity ratios, as well as uncertainty caused by the aforementioned backgrounds, overlapping events (pile-up), and, in the case of the JP sample, trigger bias studied with Monte Carlo simulation~\cite{STAR_DLL}.
%The pile-up effect was studied by examining the observed signal candidate yields for different instantaneous beam luminosities and by extrapolating these yields to vanishingly small collision rates.
%The JP trigger condition biases were studied by Monte Carlo simulation of events  generated with PYTHIA 6.4~\cite{Sjostrand:2006za} and the STAR detector response package~\cite{Geant}.
The effect of $\Lambda$ ($\bar{\Lambda}$) spin precession in the STAR magnetic field is negligible.
%The above contributions are considered to be independent and the additional details may be found in Ref.\cite{STAR_DLL}.

% \begin{wrapfigure}[29]{R}{90mm}%% [number of text lines to wrap]{horizontal position: LRC}{figure width}
% \centering
%  \vspace*{-8mm} %% the vertical position may need tweaking
%  \includegraphics[width=90mm]{figure-3}

In addition to longitudinal spin transfer, the transverse spin transfer of hyperons from proton is also of particular interest in $pp$ collisions, since it can provide access to the transverse spin content of nucleon, i.e., the transversity distribution, which is still poorly known in experiment.
Unlike the polarization is along hyperon's momentum in the case of longitudinal polarization, the azimuthal direction in the transverse plane needs to be determined first to measure transverse hyperon polarization.
E704 experiment measured the transverse spin transfer $D_{NN}$
 with respect to the production plane~\cite{Bravar:1997fb}.
% where the beam polarization needs to be projected to this direction.
Another choice is to first determine the transverse polarization direction of the fragmenting parton in a hard scattering, which is different from the polarization direction before the scattering, but can be determined by a rotation around the normal of the scattering plane~\cite{Collins:1993kq}.
The fragmenting parton's axis can be obtained via jet reconstruction with charged particle and energy deposits in calorimeters. 
Then along this direction, the transvese hyperon polarization, and thus the transverse spin transfer $D_{TT}$ can be measured.
The transverse polarization of $\Lambda$ ($\bar\Lambda$) is being investigated with hyperons reconstructed at mid-pseudorapidities with TPC at STAR.
Sizable transverse spin transfer effect is expected to exist in the large $x_F(\equiv 2p_z/{\sqrt s})$ region. 
At STAR, a Forward Hadron Calorimeter (FHC) is being proposed to be installed behind the Forward Meson Spectrometer (FMS) in the near future, which may enable the reconstruction of $\Lambda$ hyperons via the decay channel to $n\pi^0$ with $\pi^0$ detected by the FMS and $n$ by the FHC.
Simulation studies on $\Lambda$ reconstruction in the forward region using the FMS and the FHC are underway.

In summary, we made measurements on the longitudinal spin transfer to $\Lambda$ and $\bar{\Lambda}$ hyperons in $\sqrt{s} = 200\,\mathrm{GeV}$ polarized proton-proton collisions for hyperon $p_\mathrm{T}$ up to $4\,\mathrm{GeV}/c$.
The spin transfer is found to be $D_{LL}= -0.03\pm 0.13(\mathrm{stat}) \pm 0.04(\mathrm{syst})$ for $\Lambda$ and $D_{LL} = -0.12 \pm 0.08(\mathrm{stat}) \pm 0.03(\mathrm{syst})$ for $\bar{\Lambda}$ hyperons with $\left<\eta\right> = 0.5$ and $\left<p_\mathrm{T}\right> = 3.7\,\mathrm{GeV}/c$.
%The longitudinal spin transfer is sensitive to the polarized parton distribution and polarized fragmentation functions.
%The present results for $\Lambda$ and $\bar{\Lambda}$ do not provide conclusive evidence for a spin transfer signal and have uncertainties that are comparable to the variation between model expectations for the longitudinal spin transfer at RHIC.
The measurements of the transfer spin transfer may provide access to transversity distribution of the nucleon, and feasibility studies have started.

\end{document}